\documentclass[prb,twocolumn,superscriptaddress,floatfix,showkeys]{revtex4}
\usepackage{graphicx}
\usepackage{amsmath,subfigure,epsfig,psfrag}
\usepackage{comment}
\usepackage{commath}
\usepackage{graphicx}
\usepackage{amssymb}
\usepackage{xcolor}
\usepackage{soul}
\usepackage{listings}
\usepackage{rotating}
\usepackage{latexsym}
\usepackage{amsfonts}
\usepackage{amssymb}
\usepackage{wasysym}
\usepackage{xfrac}
\usepackage{mathtools}
\usepackage{multirow}
\makeatletter
\renewcommand*\env@matrix[1][\arraystretch]{%
  \edef\arraystretch{#1}%
  \hskip -\arraycolsep
  \let\@ifnextchar\new@ifnextchar
  \array{*\c@MaxMatrixCols c}}
\makeatother

\begin{document}

\title{Anharmonic Vibrational States of Double-Well Potentials in the Solid State from DFT Calculations}


\author{Davide Mitoli}
\affiliation{Dipartimento di Chimica, Universit\`{a} di Torino, via Giuria 5, 10125 Torino, Italy}

\author{Maria Petrov}
\affiliation{Dipartimento di Chimica, Universit\`{a} di Torino, via Giuria 5, 10125 Torino, Italy}

\author{Jefferson Maul}
\affiliation{Dipartimento di Chimica, Universit\`{a} di Torino, via Giuria 5, 10125 Torino, Italy}

\author{William B. Stoll}
\affiliation{Department of Chemistry, University of Rochester, Rochester, New York 14627, United States}

\author{Michael T. Ruggiero}
\email{michael.ruggiero@rochester.edu}
\affiliation{Department of Chemistry, University of Rochester, Rochester, New York 14627, United States}

\author{Alessandro Erba}
\email{alessandro.erba@unito.it}
\affiliation{Dipartimento di Chimica, Universit\`{a} di Torino, via Giuria 5, 10125 Torino, Italy}

\date{\today}

\begin{abstract}
We introduce a general approach for the simulation of quantum vibrational states of (symmetric and asymmetric) double-well potentials in molecules and materials for thermodynamic and spectroscopic applications. The method involves solving the nuclear Schr\"odinger equation associated with a one-mode potential of the type $V(Q) = aQ^2 +bQ^3 +cQ^4$ (with $a<0$ and $c>0$), and thus explicitly includes nuclear quantum effects. The potential, $V(Q)$, is obtained from density functional theory (DFT) calculations performed at displaced nuclear configurations along the selected normal mode, $Q$. The strategy has been implemented into the \textsc{Crystal} electronic structure package and allows for i) the use of many density functional approximations, including hybrid ones, and ii) integration with a quasi-harmonic module. The method is applied to the spectroscopic characterization of soft lattice modes in two phases of the molecular crystal of thiourea: a low-temperature ferroelectric phase and a high-temperature paraelectric phase. Signature peaks associated to structural changes between the two phases are found in the terahertz  region of the electromagnetic spectrum, which exhibit strong anharmonic character in their thermal evolution, as measured by temperature-dependent terahertz time-domain spectroscopy.  
\end{abstract}

\maketitle

\begin{center}
\begin{figure}[t!!]
\centering
\includegraphics[width=7.2cm]{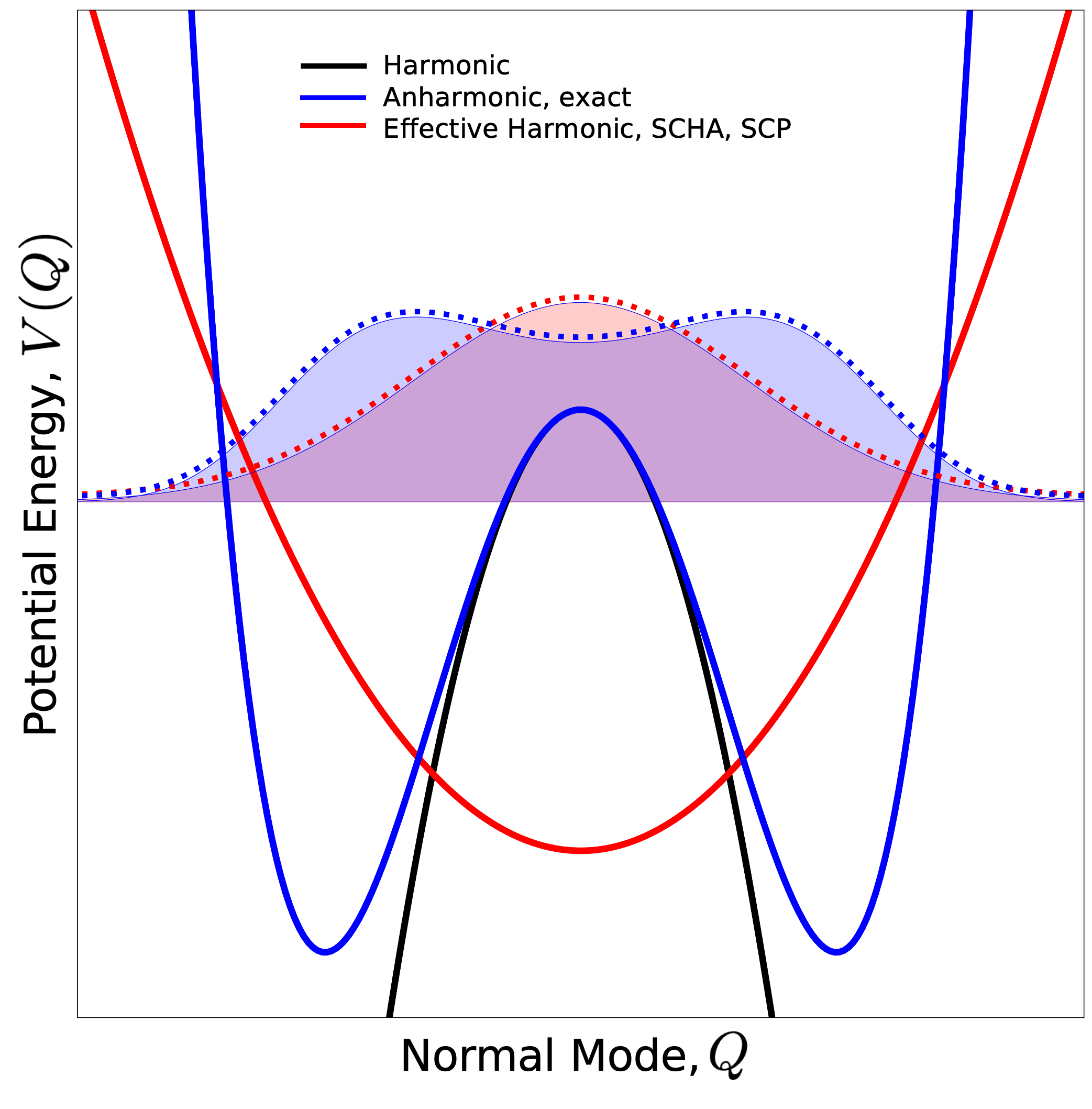}
\caption{Schematic representation of a strongly anharmonic, symmetric, double-well potential (blue solid line), along with an optimized effective harmonic potential that approximates it within the SCHA and SCP theories (red solid line). The corresponding lowest-energy eigenstates are shown with dashed blue and red lines, respectively. The actual harmonic term of the true potential is shown with a black solid line.}
\label{fig:1}
\end{figure}
\end{center}

An explicit treatment of nuclear degrees of freedom through statistical mechanics is key to a reliable description of finite temperature properties of materials: stability, structure, phase transitions, transport, and so on. With first-principles simulations based on the Born-Oppenheimer (BO) approximation, lattice dynamics is routinely described within the harmonic approximation (HA), which is valid only for weakly anharmonic systems and for restricted temperature ranges. Moreover, many properties of materials are intrinsically anharmonic and thus can not be captured by the HA (e.g. lattice thermal conductivity, phonon lifetimes, thermal expansion, pyroelectricity).\cite{garg2011role,hellman2011lattice,errea2013first,bansal2016phonon,CU2O_ANTTI,zhou2014lattice,tadano2014anharmonic,rossi2016anharmonic,eklund2023pyroelectric,liu2018mechanisms,PURINE_QHA,JunejaThermalExp} For a given BO potential (potential energy surface, PES), exact anharmonic free energies can be calculated from imaginary time path integral simulations.\cite{kapil2019assessment} However, their prohibitive computational cost has so far prevented their popularity in favor of other, more efficient, approximate approaches, among which the self-consistent harmonic approximation (SCHA) and self-consistent phonon (SCP) theory have become the standard in solid state physics.\cite{hooton1955li,werthamer1970self,PhysRevLett.100.095901,errea2014anharmonic,monacelli2021stochastic,PhysRevB.103.104305,PhysRevB.107.174307,tadano2015self,zacharias2023anharmonic,schiltz2023implementation,monacelli2024simulating} Within such approaches, anharmonicity is described through an effective temperature-dependent harmonic Hamiltonian with the assumption of Gaussian atomic fluctuations (see Figure \ref{fig:1} for a schematic graphical representation).\cite{errea2014anharmonic}

Among strongly anharmonic potentials, the double-well potential (DWP) is ubiquitous in solid state physics as it plays a central role in structural phase transitions, ferroelectricity, proton-transfers, and in general in finite-temperature structure and stability.\cite{zacharias2023anharmonic,tadano2019ab,hoffmann2019unveiling,konwent1986application,choudhury2003role,wang2002quantum,fillaux1998new,fillaux2002impact,xu2017vibrational,krasilnikov2014two,eckold1992proton,cailleau1980double,goryainov2012model} The Gaussian approximation of SCHA and SCP breaks down in this case as quantum tunneling in multi-minima energy landscapes yields wavefunctions that significantly depart from a Gaussian behavior, and instead often exhibit multi-peak shapes;\cite{siciliano2024beyond} an approach to partially correct for this limitation within the SCHA (namely, non-linear SCHA) has very recently been formulated.\cite{siciliano2024beyond} Figure \ref{fig:1} reports a schematic representation of a strongly anharmonic DWP and of its associated lowest-energy eigenstate (blue lines), and a comparison with an effective harmonic potential and corresponding eigenstate that approximates it within the SCHA and SCP approaches (red lines). In this Letter, we present a general method to obtain both energy levels and wavefunctions of vibrational states of symmetric and asymmetric DWPs in solids. Solutions of the nuclear Schr\"odinger equation are found, associated with an anharmonic PES derived from density functional theory (DFT) calculations. 
The method allows for fully including the true anharmonic character of vibrational potentials to compute anharmonic wavefunctions, leading to an accurate description of properties that depend explicitly on anharmonic states, such as spectroscopic transitions, thermal displacement parameters, pyroelectricity, and so on. 

The proposed strategy has been implemented in a developer's version of the \textsc{Crystal23} electronic structure package\cite{erba2022crystal23} and represents an extension of the anharmonic module thereof for: i) calculation of cubic and quartic interatomic force constants;\cite{PARTI_ANHARM, mitoli2023anharmonic} ii) non-perturbative many-body solution of the nuclear Schr\"odinger equation via the vibrational self-consistent field (VSCF) and vibrational configuration interaction (VCI) methods;\cite{PARTII_ANHARM,maul2019elucidating,schireman2022anharmonic} iii) infrared and Raman anharmonic vibrational spectroscopy.\cite{carbonniere2020calculation,mitoli2024first}

The method starts with the calculation of phonon frequencies $\omega_i$ and normal modes $Q_i$ (with $i=1,\dots,3N$, where $N$ represents the number of atoms per cell) from the HA. Normal mode coordinates associated to DWPs are identified from imaginary frequencies. We work within a quartic representation of the DWP as follows:
\begin{equation}
\label{eq:DWP}
V = aQ^2 +bQ^3 +cQ^4   \; ,
\end{equation}
with $a<0$ and $c>0$. The coefficient $b$ of the cubic term determines the skewness of the potential (symmetric for $b=0$, asymmetric for $b\neq 0$). The accurate numerical evaluation of the $a$, $b$ and $c$ parameters of the DWP of Eq. (\ref{eq:DWP}) requires some care, especially when the wells are shallow. For instance, they can be computed in terms of second-, third- and fourth-order energy derivatives with respect to $Q$ via finite-difference approaches, see Eq. (S1) in the Supporting Information. Three different finite-difference approaches have been implemented in \textsc{Crystal} for evaluation of cubic and quartic interatomic force constants, some based only on the energy and others based on both the energy and the (analytical) forces.\cite{PARTI_ANHARM} The most efficient approach belongs to the latter class, and requires calculations at just two displaced nuclear configurations at $\pm \delta$ along $Q$.\cite{PARTI_ANHARM,lin2008calculating} With this approach, we find the quartic coefficient $c$ to significantly depend on the step $\delta$. In particular, a larger step than usually set for less anharmonic potentials provides better results ($\delta = 1.5$ relative to the default value of 0.9, in units of classical amplitude). An alternative approach, based on a denser and more extended sampling of the PES, consists in computing the energy values at several displaced nuclear configurations along the normal mode and best-fit them to a polynomial representation of $V(Q)$ as in Eq. (\ref{eq:DWP}) to obtain the $a$, $b$ and $c$ parameters. Here, the most critical factor is the explored $Q$ range. In general this approach provides more accurate results at a higher cost. We analyze the numerical performance of these approaches in the Supporting Information.

\begin{center}
\begin{figure}[t!!]
\centering
\includegraphics[width=8.6cm]{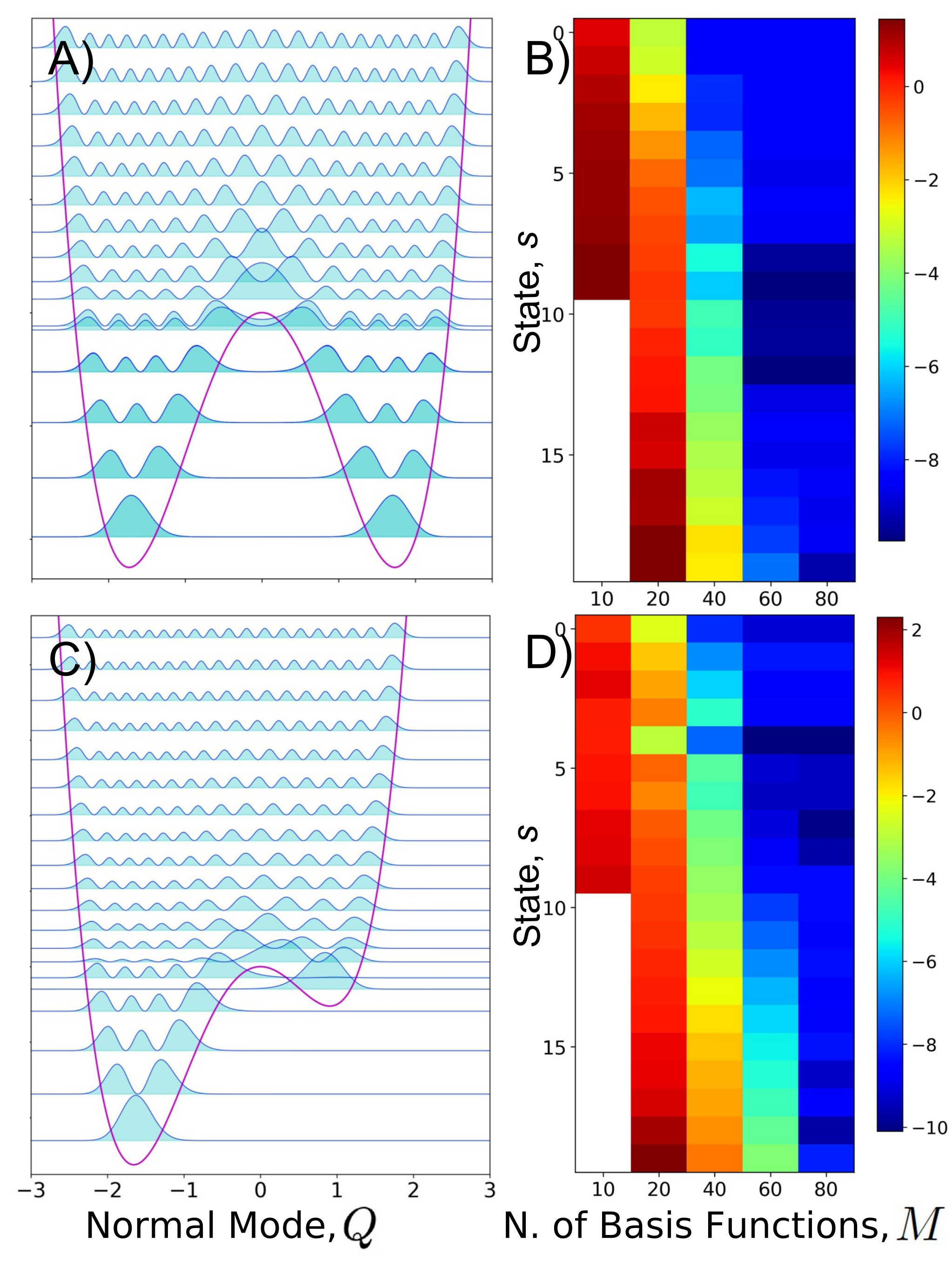}
\caption{A) The 20 lowest lying anharmonic states (vibrational energy levels and square-modulus of the corresponding wavefunctions) of a symmetric DWP with $a=-30$ and $c=5$. B) Convergence of the 20 lowest lying anharmonic states of the DWP represented in panel A) as a function of the number of basis functions used, $M$; for each state $s$, the reported quantity is log$_{10}[\vert E_s(M) - E_s(M\to \infty)\vert ]$. C) Same as in A) but for an asymmetric DWP with $a=-30$, $b=10$ and $c=10$. D) Same as in B) but for the asymmetric potential of panel C). Plots in panels A) and C) are produced with the CRYSTALpytools Python interface to \textsc{Crystal}.\cite{CRYSTALpytools}}
\label{fig:2}
\end{figure}
\end{center}

\noindent We consider a one-mode nuclear Hamiltonian:
\begin{equation}
\label{eq:nucH}
H =  -\frac{1}{2}\frac{\partial^2}{\partial Q^2} + V(Q) \; ,
\end{equation}
where $V(Q)$ is the DWP of Eq. (\ref{eq:DWP}), and look for its solutions $H \Psi_s = E_s \Psi_s$ (with $s$ being a state index). We express the anharmonic wavefunctions $\Psi_s$ in an appropriately selected (see below) harmonic basis:
\begin{equation}
\label{eq:psi}
\Psi_s(Q) = \sum_{\mu =1}^M c_{\mu,s} \psi_\mu(Q;\alpha) \; ,
\end{equation}
where $\psi_\mu(Q;\alpha)$ are the solutions of a quantum harmonic oscillator with potential $\frac{1}{2} \alpha^4 Q^2$:
\begin{equation} 
\label{eq:HAwf2}
\psi_{\mu}(Q;\alpha) = \left( \frac{\alpha}{\sqrt{\pi}2^\mu \mu !} \right) ^{\frac{1}{2}} H_{\mu}(\xi) e^{- \frac{\xi^2}{2}} \; ,
\end{equation}
with $\xi = Q \alpha$ and $H_\mu$ being the $\mu$-th order Hermite polynomial. To determine the optimal harmonic basis (i.e. optimal value of the parameter $\alpha$) for the description of the DWP of Eq. (\ref{eq:DWP}), we follow the strategy first suggested by Balsa {\it et al.} and set $\alpha^2 = \sqrt{\vert a\vert}$.\cite{balsa1983simple} Upon linearization as in Eq. (\ref{eq:psi}), the nuclear Schr\"odinger equation can be expressed in matrix form,
${\bf H} {\bf C} = {\bf E} {\bf C}$, and its solutions obtained by diagonalization of the symmetric Hamiltonian matrix ${\bf H}$, whose non-vanishing elements have closed analytical expressions given below:
\begin{eqnarray} 
\langle{\mu} \vert H\vert {\mu} \rangle
    &\equiv& \langle{\mu} \vert T +aQ^2 +cQ^4\vert {\mu} \rangle \nonumber \\
    &=& \left (\frac{\alpha^4 + 2a}{2\alpha^2} \right) 
    \left(\mu +\frac{1}{2} \right) \nonumber \\
    &+& \frac{3c}{4\alpha^4} (2\mu^2 + 2\mu + 1)\nonumber \\
\langle{\mu} \vert H \vert {\mu+1} \rangle &\equiv& \langle{\mu} \vert bQ^3\vert {\mu+1} \rangle \nonumber \\
    &=& \frac{3b}{2} \sqrt{\frac{(\mu + 1)^3} {2\alpha^6}}  \nonumber \\
\langle{\mu} \vert H \vert {\mu+2} \rangle
    &\equiv& \langle{\mu} \vert T+aQ^2 +cQ^4\vert {\mu+2} \rangle \nonumber \\
    &=& \sqrt{(\mu+1)(\mu+2)}\nonumber \\
    &\times&\left[ -\frac{\alpha^2}{4}+\frac{a}{2\alpha^2}+ \frac{c}{2\alpha^4}(2\mu+3)\right] \nonumber \\
\langle{\mu} \vert H \vert {\mu+3} \rangle &\equiv& \langle{\mu} \vert bQ^3\vert {\mu+3} \rangle \nonumber \\
    &=& b\sqrt{\frac{(\mu + 1) (\mu + 2) (\mu+3)} {8\alpha^6}}  \nonumber \\
\langle{\mu} \vert H \vert {\mu+4} \rangle 
    &\equiv& \langle{\mu} \vert cQ^4\vert {\mu+4} \rangle \nonumber \\
    &=& \frac{c}{4\alpha^4}
    \sqrt{(\mu+1)(\mu+2)(\mu+3)}  \; ,
\end{eqnarray}
where $T$ represents the kinetic term of the nuclear Hamiltonian. It is crucial to check for convergence of the obtained anharmonic states with respect to the size of the adopted harmonic basis -- i.e. with respect to the number $M$ of basis functions used in Eq. (\ref{eq:psi}) -- both in terms of energies $E_s$ and wavefunctions $\Psi_s$. We address this aspect in Figure \ref{fig:2} for symmetric and asymmetric DWPs. In both cases, we analyze the 20 lowest lying anharmonic states. Panels A) and C) report a graphical representation of the fully converged ($M\to \infty$, i.e. $M = 500$ in this case) exact solutions in terms of vibrational energy levels and square-modulus of the corresponding wavefunctions. Panels B) and D) show the convergence of the anharmonic energy levels $E_s$ with respect to $M$. The reported quantity, for each state $s$, is log$_{10}[\vert E_s(M) - E_s(M\to \infty)\vert ]$, which is given on a color scale. The following can be observed: i) the computed anharmonic states regularly converge as $M$ increases (for both potentials, $M\geq 60$ provides stable results); ii) the lowest energy states converge with $M$ faster than the higher energy ones; iii) the convergence with $M$ is faster for symmetric rather than asymmetric potentials.

We apply the methodology described above to the molecular crystal of thiourea to document its effectiveness. Thiourea has a rich polymorphism as a function of temperature, pressure, applied electric field, and is the second oldest ferroelectric to exhibit an incommensurate phase, between 169 and 202 K.\cite{mckenzie1973dielectric,klimowski1976effect,dove1986model} Below 169 K, thiourea crystallizes in a ferroelectric orthorhombic $P2_1ma$ lattice (low-temperature phase). Above 202 K, a paraelectric orthorhombic phase is stabilized with a lattice of $Pnma$ space group symmetry (high-temperature phase). These two phases are very similar, with four molecules per unit cell and a structure consisting of staggered ribbons along the {\bf b} axis. The molecules are polar. In the high-temperature, paraelectric, phase, the dipole moments of two pairs of adjacent chains cancel out due to symmetry, resulting in no net dipole moment. On the other hand in the low-temperature, ferroelectric, phase, the molecules slightly tilt and shift in the {\bf ac} plane (compared to the high-temperature phase), resulting in a non-vanishing net polarization.\cite{elcombe1968neutron} Figure \ref{fig:3} A) and B) reports a view down the {\bf b} axis of the crystal structure of these two phases of thiourea. The ferroelectric and incommensurate phase transitions only involve molecular motions in the {\bf ac} plane. 
Soft lattice vibrations associated with such motions were probed by vibrational spectroscopies (infrared reflection and Raman scattering) in the 1970s and were found to produce peaks in the terahertz region of the spectrum.\cite{siapkas1980soft,petzelt1981dielectric} In particular, their temperature dependence was analyzed in the temperature region 20-300 K, which showed remarkable anharmonic features, including the pronounced broadening of the lowest active infrared peak. Here, we find this peak to be associated to a DWP in the high-temperature paraelectric phase, which explains the observed broadening (see below).  

Experimental terahertz time-domain spectroscopy measurements were performed on commercially-acquired microcrystalline powders of thiourea (Sigma Aldrich, $>99\%$). The sample was used as-received, mixed with polytetrafluoroethylene (PTFE) to a $1\%$ w/w concentration, homogenized with a mortar and pestle, and pressed into 13 mm diameter free-standing pellets with a width of ca. 3 mm. A pellet of pure PTFE with the same dimensions was created and served as a spectral reference. The samples were placed in a liquid nitrogen cryostat (Lakeshore Cryotronics), which permitted acquisition at temperatures from 78 K to 300 K. The terahertz spectra were acquired using a commercial terahertz spectrometer (Toptica Photonics), which consisted of a pair of fiber-coupled emitter and receiver modules, and a corresponding free-space optical setup for collimating and focusing the terahertz radiation on the sample and receiver. The recorded terahertz time-domain waveforms were a result of 20,000 individual averaged datasets, which were subsequently Fourier transformed to yield frequency-domain power spectra. An absorption spectrum was generating by taking the ratio of a sample and blank power spectrum, and the  spectra shown in Figure \ref{fig:3} E) and F) (dashed lines) are a result of averaging four individual absorption spectra in the frequency domain. The experimental spectra were acquired at select temperatures: 78 K for the ferroelectric P$2_1ma$ phase and 200, 250 and 300 K for the paraelectric P$nma$ phase. Two narrow peaks are present at low temperature in the ferroelectric phase at about 30 and 55 cm$^{-1}$. The second peak disappears in the high-temperature paraelectric phase while the first one broadens very significantly; as temperature increases from 200 to 300 K, the first peak slightly shifts to higher wavenumbers and progressively broadens.

We perform DFT calculations with the \textsc{Crystal} electronic structure package\cite{erba2022crystal23,MPP2017} on the low- and high-temperature phases of thiourea with the hybrid B3LYP exchange-correlation functional,\cite{B3LYP} as corrected for missing dispersive interactions following Grimme's -D3 approach.\cite{GRIMMED3} A triple-zeta quality basis set specifically optimized for solid state calculations is used.\cite{vilela2019bsse} Full structural relaxation is achieved through a geometry optimization process, followed by the evaluation of harmonic frequencies and normal modes of vibration. All computed harmonic frequencies are positive in the low-temperature P$2_1ma$ structure, while the high-temperature P$nma$ structure exhibits one negative eigenvalue of the mass-weighted Hessian, which could be indicative of a DWP. We determine the coefficients of the DWP in the paraelectric phase of thiourea by computing the DFT energy at 11 displaced nuclear configurations along the corresponding normal mode and by fitting them to the potential in Eq. (\ref{eq:DWP}), see blue circles and line in Figure \ref{fig:3} C). The solutions -- energy levels and wavefunctions -- of the associated nuclear Schr\"odinger equation are also shown in Figure \ref{fig:3} C), as obtained following the approach outlined above, with $M=150$, which ensures full convergence of all spectroscopically relevant states.

\begin{widetext}
\begin{center}
\begin{figure}[h!!]
\centering
\includegraphics[width=17.8cm]{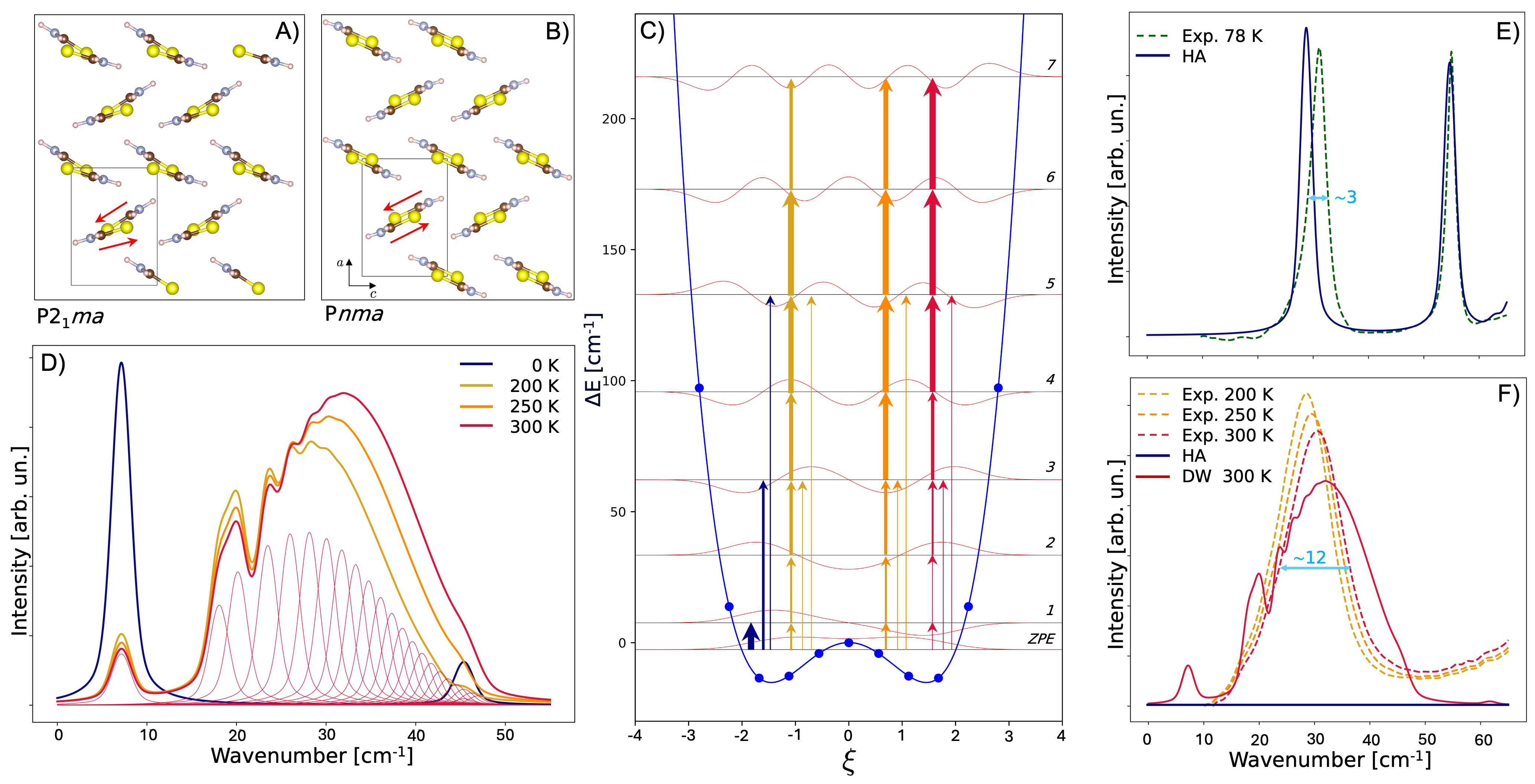}
\caption{A) and B) Crystal structure of the low-temperature ferroelectric P$2_1ma$ and high-temperature paraelectric P$nma$ orthorhombic phase of thiourea, respectively (viewed down the {\bf b} axis). C) The DWP of the P$nma$ phase (blue circles and line) as derived from DFT calculations, along with the  solutions -- energy levels and wavefunctions -- to its associated nuclear Schr\"odinger equation. The arrows mark active infrared transitions with intensities proportional to their thickness (dark blue, 0 K, yellow, 200 K, orange, 250 K, red, 300 K). D) Simulated infrared spectrum associated to the DWP of panel C) at four different temperatures. Single transitions contributing to the overall spectrum at 300 K are reported as thin red lines. E) Infrared spectrum of the low-temperature ferroelectric phase in the 0-65 cm$^{-1}$ spectral range (dashed line for experiment at 78 K, blue solid line for the HA simulation). F) Infrared spectrum of the high-temperature paraelectric phase in the 0-65 cm$^{-1}$ spectral range (dashed lines for experiment at 200, 250 and 300 K, blue solid line for the HA simulation, red solid line for the anharmonic simulation). All simulated spectra are scaled by 0.7 to match the experimental peak positions.}
\label{fig:3}
\end{figure}
\end{center}
\end{widetext}

Key to a correct description of the spectroscopic fingerprint of a DWP is the description of anharmonic ``hot bands'' (i.e. spectral features associated to transitions between two vibrationally excited states). Infrared (IR) intensities associated to a transition between an initial state ${\Psi_i}$, with energy $E_i$, and a final state ${\Psi_f}$ can be modeled as:
\begin{equation}
\label{eq:int00}
I^{\text{IR}}_{i\rightarrow f} (T) = I^{\text{IR},0}_{i\rightarrow f} \times p_i (T) \; ,
\end{equation}
By expanding the dipole moment ${\boldsymbol \mu}$ in a Taylor series with respect to $Q$ and truncating to first-order (i.e. neglecting electrical anharmonicity), the first term on the right-hand side of Eq. (\ref{eq:int00}) reads:
\begin{equation}
\label{eq:IR_intgen}
I^{\text{IR},0}_{i\rightarrow f} = \sum_{a=x,y,z} 
\left( \frac{\partial \mu_a}{\partial Q}\right)^2
\langle i | Q | f\rangle^2 \; ,
\end{equation}
where the derivatives constitute the Born tensor that we compute either numerically through a Berry phase approach or analytically through a coupled-perturbed (CP) approach.\cite{maschio2012ab,dovesi2018calculation} The last term in Eq. (\ref{eq:int00}) represents the statistical probability of the initial state to be thermally populated:
\begin{equation}
p_i (T) = \frac{1}{Z} e^{-\frac{E_i} {k_B T}} \quad \textup{with} \quad Z = \sum_{s}^\textup{all} e^{-\frac{E_s} {k_B T}} \; .
\end{equation}
The main active IR transitions associated to the DWP of thiourea are schematically shown in Figure \ref{fig:3} C) at four selected temperatures (0, 200, 250 and 300 K) by arrows with different thickness (proportional to the corresponding intensity). At 0 K only transitions starting from the fundamental state would be described (dark blue arrows). At higher temperatures, a multitude of ``hot band'' transitions occur, which determine the overall broadening of the DW spectral feature in the infrared spectrum. This is shown in Figure \ref{fig:3} D) where we report the simulated IR spectrum of the paraelectric phase of thiourea at four temperatures. Neglecting thermal effects (i.e. no ``hot bands'' at 0 K) results in a narrow peak at about 7 cm$^{-1}$. Simulations at finite temperatures corresponding to the stability domain of the paraelectric phase, show several distinctive features: i) a broad spectral feature centered at about 30 cm$^{-1}$; ii) a slight shift of the peak to higher wavenumbers as a function of temperature; iii) a slight increase in the broadening as a function of temperature. For the 300 K case, red lines, we also report the individual transitions contributing to the total profile to highlight the role of the multiple ``hot bands'' in determining the shape of the spectral feature of a DWP.        

We compare the simulated spectra to the experimental ones in the 0-65 cm$^{-1}$ spectral range in Figure \ref{fig:3} E) and F). In particular, we show the simulated IR spectrum for the low-temperature ferroelectric phase as obtained from the HA (blue solid line) in panel E). The HA completely breaks down in the high-temperature paraelectric phase, where it predicts no peaks in the whole 0-65 cm$^{-1}$ range, see blue solid line in panel F). The simulated spectrum from the anharmonic approach of the DWP discussed above is shown as a solid red line, as modeled at 300 K, which matches rather remarkably with the experiment.      

In conclusion, we have presented a theoretical method and a computational protocol to effectively describe vibrational states associated to strongly anharmonic, double-well, potentials within DFT calculations. The effectiveness of the presented approach has been discussed on a DWP found in the high-temperature paraelectric phase of the thiourea molecular crystal. Work is currently in progress to extend this approach to couplings of the DWP with other modes.

\acknowledgments

This research has received funding from the Project CH4.0 under the MUR program ``Dipartimenti di Eccellenza 2023-2027'' (CUP: D13C22003520001). MTR and WBS acknowledge support from the National Science Foundation (award numbers CHE-2055417 and DMR-2348765).

\bibliographystyle{achemso}
\bibliography{DatabaseBIBLIO}

\providecommand{\latin}[1]{#1}
\makeatletter
\providecommand{\doi}
  {\begingroup\let\do\@makeother\dospecials
  \catcode`\{=1 \catcode`\}=2 \doi@aux}
\providecommand{\doi@aux}[1]{\endgroup\texttt{#1}}
\makeatother
\providecommand*\mcitethebibliography{\thebibliography}
\csname @ifundefined\endcsname{endmcitethebibliography}  {\let\endmcitethebibliography\endthebibliography}{}
\begin{mcitethebibliography}{60}
\providecommand*\natexlab[1]{#1}
\providecommand*\mciteSetBstSublistMode[1]{}
\providecommand*\mciteSetBstMaxWidthForm[2]{}
\providecommand*\mciteBstWouldAddEndPuncttrue
  {\def\EndOfBibitem{\unskip.}}
\providecommand*\mciteBstWouldAddEndPunctfalse
  {\let\EndOfBibitem\relax}
\providecommand*\mciteSetBstMidEndSepPunct[3]{}
\providecommand*\mciteSetBstSublistLabelBeginEnd[3]{}
\providecommand*\EndOfBibitem{}
\mciteSetBstSublistMode{f}
\mciteSetBstMaxWidthForm{subitem}{(\alph{mcitesubitemcount})}
\mciteSetBstSublistLabelBeginEnd
  {\mcitemaxwidthsubitemform\space}
  {\relax}
  {\relax}

\bibitem[Garg \latin{et~al.}(2011)Garg, Bonini, Kozinsky, and Marzari]{garg2011role}
Garg,~J.; Bonini,~N.; Kozinsky,~B.; Marzari,~N. Role of disorder and anharmonicity in the thermal conductivity of silicon-germanium alloys: A first-principles study. \emph{Phys. Rev. Lett.} \textbf{2011}, \emph{106}, 045901\relax
\mciteBstWouldAddEndPuncttrue
\mciteSetBstMidEndSepPunct{\mcitedefaultmidpunct}
{\mcitedefaultendpunct}{\mcitedefaultseppunct}\relax
\EndOfBibitem
\bibitem[Hellman \latin{et~al.}(2011)Hellman, Abrikosov, and Simak]{hellman2011lattice}
Hellman,~O.; Abrikosov,~I.~A.; Simak,~S.~I. Lattice dynamics of anharmonic solids from first principles. \emph{Phys. Rev. B} \textbf{2011}, \emph{84}, 180301\relax
\mciteBstWouldAddEndPuncttrue
\mciteSetBstMidEndSepPunct{\mcitedefaultmidpunct}
{\mcitedefaultendpunct}{\mcitedefaultseppunct}\relax
\EndOfBibitem
\bibitem[Errea \latin{et~al.}(2013)Errea, Calandra, and Mauri]{errea2013first}
Errea,~I.; Calandra,~M.; Mauri,~F. First-principles theory of anharmonicity and the inverse isotope effect in superconducting palladium-hydride compounds. \emph{Phys. Rev. Lett.} \textbf{2013}, \emph{111}, 177002\relax
\mciteBstWouldAddEndPuncttrue
\mciteSetBstMidEndSepPunct{\mcitedefaultmidpunct}
{\mcitedefaultendpunct}{\mcitedefaultseppunct}\relax
\EndOfBibitem
\bibitem[Bansal \latin{et~al.}(2016)Bansal, Hong, Li, May, Porter, Hu, Abernathy, and Delaire]{bansal2016phonon}
Bansal,~D.; Hong,~J.; Li,~C.~W.; May,~A.~F.; Porter,~W.; Hu,~M.~Y.; Abernathy,~D.~L.; Delaire,~O. Phonon anharmonicity and negative thermal expansion in SnSe. \emph{Phys. Rev. B} \textbf{2016}, \emph{94}, 054307\relax
\mciteBstWouldAddEndPuncttrue
\mciteSetBstMidEndSepPunct{\mcitedefaultmidpunct}
{\mcitedefaultendpunct}{\mcitedefaultseppunct}\relax
\EndOfBibitem
\bibitem[Linnera \latin{et~al.}(2019)Linnera, Erba, and Karttunen]{CU2O_ANTTI}
Linnera,~J.; Erba,~A.; Karttunen,~A.~J. Negative thermal expansion of Cu$_2$O studied by quasi-harmonic approximation and cubic force-constant method. \emph{J. Chem. Phys.} \textbf{2019}, \emph{151}, 184109\relax
\mciteBstWouldAddEndPuncttrue
\mciteSetBstMidEndSepPunct{\mcitedefaultmidpunct}
{\mcitedefaultendpunct}{\mcitedefaultseppunct}\relax
\EndOfBibitem
\bibitem[Zhou \latin{et~al.}(2014)Zhou, Nielson, Xia, Ozoli{\c{n}}{\v{s}}, \latin{et~al.} others]{zhou2014lattice}
Zhou,~F.; Nielson,~W.; Xia,~Y.; Ozoli{\c{n}}{\v{s}},~V.; others Lattice anharmonicity and thermal conductivity from compressive sensing of first-principles calculations. \emph{Phys. Rev. Lett.} \textbf{2014}, \emph{113}, 185501\relax
\mciteBstWouldAddEndPuncttrue
\mciteSetBstMidEndSepPunct{\mcitedefaultmidpunct}
{\mcitedefaultendpunct}{\mcitedefaultseppunct}\relax
\EndOfBibitem
\bibitem[Tadano \latin{et~al.}(2014)Tadano, Gohda, and Tsuneyuki]{tadano2014anharmonic}
Tadano,~T.; Gohda,~Y.; Tsuneyuki,~S. Anharmonic force constants extracted from first-principles molecular dynamics: applications to heat transfer simulations. \emph{J. Phys. Condens. Matter} \textbf{2014}, \emph{26}, 225402\relax
\mciteBstWouldAddEndPuncttrue
\mciteSetBstMidEndSepPunct{\mcitedefaultmidpunct}
{\mcitedefaultendpunct}{\mcitedefaultseppunct}\relax
\EndOfBibitem
\bibitem[Rossi \latin{et~al.}(2016)Rossi, Gasparotto, and Ceriotti]{rossi2016anharmonic}
Rossi,~M.; Gasparotto,~P.; Ceriotti,~M. Anharmonic and quantum fluctuations in molecular crystals: A first-principles study of the stability of paracetamol. \emph{Phys. Rev. Lett.} \textbf{2016}, \emph{117}, 115702\relax
\mciteBstWouldAddEndPuncttrue
\mciteSetBstMidEndSepPunct{\mcitedefaultmidpunct}
{\mcitedefaultendpunct}{\mcitedefaultseppunct}\relax
\EndOfBibitem
\bibitem[Eklund and Karttunen(2023)Eklund, and Karttunen]{eklund2023pyroelectric}
Eklund,~K.; Karttunen,~A.~J. Pyroelectric Effect in Tetragonal Ferroelectrics BaTiO3 and KNbO3 Studied with Density Functional Theory. \emph{J. Phys. Chem. C} \textbf{2023}, \emph{127}, 21806--21815\relax
\mciteBstWouldAddEndPuncttrue
\mciteSetBstMidEndSepPunct{\mcitedefaultmidpunct}
{\mcitedefaultendpunct}{\mcitedefaultseppunct}\relax
\EndOfBibitem
\bibitem[Liu and Pantelides(2018)Liu, and Pantelides]{liu2018mechanisms}
Liu,~J.; Pantelides,~S.~T. Mechanisms of pyroelectricity in three-and two-dimensional materials. \emph{Phys. Rev. Lett.} \textbf{2018}, \emph{120}, 207602\relax
\mciteBstWouldAddEndPuncttrue
\mciteSetBstMidEndSepPunct{\mcitedefaultmidpunct}
{\mcitedefaultendpunct}{\mcitedefaultseppunct}\relax
\EndOfBibitem
\bibitem[Ruggiero \latin{et~al.}(2017)Ruggiero, Zeitler, and Erba]{PURINE_QHA}
Ruggiero,~M.~T.; Zeitler,~J.; Erba,~A. Intermolecular Anharmonicity in Molecular Crystals: Interplay between Experimental Low-Frequency Dynamics and Quantum Quasi-Harmonic Simulations of Solid Purine. \emph{Chem. Commun.} \textbf{2017}, \emph{53}, 3781--3784\relax
\mciteBstWouldAddEndPuncttrue
\mciteSetBstMidEndSepPunct{\mcitedefaultmidpunct}
{\mcitedefaultendpunct}{\mcitedefaultseppunct}\relax
\EndOfBibitem
\bibitem[Juneja \latin{et~al.}(2024)Juneja, Hastings, Stoll, Brennessel, Zarrella, Sornberger, Catalano, Korter, and Ruggiero]{JunejaThermalExp}
Juneja,~N.; Hastings,~J.~L.; Stoll,~W.~B.; Brennessel,~W.~W.; Zarrella,~S.; Sornberger,~P.; Catalano,~L.; Korter,~T.~M.; Ruggiero,~M.~T. Fundamentally intertwined: anharmonic intermolecular interactions dictate both thermal expansion and terahertz lattice dynamics in molecular crystals. \emph{Chem. Commun.} \textbf{2024}, --\relax
\mciteBstWouldAddEndPuncttrue
\mciteSetBstMidEndSepPunct{\mcitedefaultmidpunct}
{\mcitedefaultendpunct}{\mcitedefaultseppunct}\relax
\EndOfBibitem
\bibitem[Kapil \latin{et~al.}(2019)Kapil, Engel, Rossi, and Ceriotti]{kapil2019assessment}
Kapil,~V.; Engel,~E.; Rossi,~M.; Ceriotti,~M. Assessment of approximate methods for anharmonic free energies. \emph{J. Chem. Theory Comput.} \textbf{2019}, \emph{15}, 5845--5857\relax
\mciteBstWouldAddEndPuncttrue
\mciteSetBstMidEndSepPunct{\mcitedefaultmidpunct}
{\mcitedefaultendpunct}{\mcitedefaultseppunct}\relax
\EndOfBibitem
\bibitem[Hooton(1955)]{hooton1955li}
Hooton,~D. LI. A new treatment of anharmonicity in lattice thermodynamics: I. \emph{London Edinburgh Philos. Mag. \& J. Sci.} \textbf{1955}, \emph{46}, 422--432\relax
\mciteBstWouldAddEndPuncttrue
\mciteSetBstMidEndSepPunct{\mcitedefaultmidpunct}
{\mcitedefaultendpunct}{\mcitedefaultseppunct}\relax
\EndOfBibitem
\bibitem[Werthamer(1970)]{werthamer1970self}
Werthamer,~N. Self-consistent phonon formulation of anharmonic lattice dynamics. \emph{Phys. Rev. B} \textbf{1970}, \emph{1}, 572\relax
\mciteBstWouldAddEndPuncttrue
\mciteSetBstMidEndSepPunct{\mcitedefaultmidpunct}
{\mcitedefaultendpunct}{\mcitedefaultseppunct}\relax
\EndOfBibitem
\bibitem[Souvatzis \latin{et~al.}(2008)Souvatzis, Eriksson, Katsnelson, and Rudin]{PhysRevLett.100.095901}
Souvatzis,~P.; Eriksson,~O.; Katsnelson,~M.~I.; Rudin,~S.~P. Entropy Driven Stabilization of Energetically Unstable Crystal Structures Explained from First Principles Theory. \emph{Phys. Rev. Lett.} \textbf{2008}, \emph{100}, 095901\relax
\mciteBstWouldAddEndPuncttrue
\mciteSetBstMidEndSepPunct{\mcitedefaultmidpunct}
{\mcitedefaultendpunct}{\mcitedefaultseppunct}\relax
\EndOfBibitem
\bibitem[Errea \latin{et~al.}(2014)Errea, Calandra, and Mauri]{errea2014anharmonic}
Errea,~I.; Calandra,~M.; Mauri,~F. Anharmonic free energies and phonon dispersions from the stochastic self-consistent harmonic approximation: Application to platinum and palladium hydrides. \emph{Phys. Rev. B} \textbf{2014}, \emph{89}, 064302\relax
\mciteBstWouldAddEndPuncttrue
\mciteSetBstMidEndSepPunct{\mcitedefaultmidpunct}
{\mcitedefaultendpunct}{\mcitedefaultseppunct}\relax
\EndOfBibitem
\bibitem[Monacelli \latin{et~al.}(2021)Monacelli, Bianco, Cherubini, Calandra, Errea, and Mauri]{monacelli2021stochastic}
Monacelli,~L.; Bianco,~R.; Cherubini,~M.; Calandra,~M.; Errea,~I.; Mauri,~F. The stochastic self-consistent harmonic approximation: calculating vibrational properties of materials with full quantum and anharmonic effects. \emph{J. Phys. Condens. Matter} \textbf{2021}, \emph{33}, 363001\relax
\mciteBstWouldAddEndPuncttrue
\mciteSetBstMidEndSepPunct{\mcitedefaultmidpunct}
{\mcitedefaultendpunct}{\mcitedefaultseppunct}\relax
\EndOfBibitem
\bibitem[Monacelli and Mauri(2021)Monacelli, and Mauri]{PhysRevB.103.104305}
Monacelli,~L.; Mauri,~F. Time-dependent self-consistent harmonic approximation: Anharmonic nuclear quantum dynamics and time correlation functions. \emph{Phys. Rev. B} \textbf{2021}, \emph{103}, 104305\relax
\mciteBstWouldAddEndPuncttrue
\mciteSetBstMidEndSepPunct{\mcitedefaultmidpunct}
{\mcitedefaultendpunct}{\mcitedefaultseppunct}\relax
\EndOfBibitem
\bibitem[Siciliano \latin{et~al.}(2023)Siciliano, Monacelli, Caldarelli, and Mauri]{PhysRevB.107.174307}
Siciliano,~A.; Monacelli,~L.; Caldarelli,~G.; Mauri,~F. Wigner Gaussian dynamics: Simulating the anharmonic and quantum ionic motion. \emph{Phys. Rev. B} \textbf{2023}, \emph{107}, 174307\relax
\mciteBstWouldAddEndPuncttrue
\mciteSetBstMidEndSepPunct{\mcitedefaultmidpunct}
{\mcitedefaultendpunct}{\mcitedefaultseppunct}\relax
\EndOfBibitem
\bibitem[Tadano and Tsuneyuki(2015)Tadano, and Tsuneyuki]{tadano2015self}
Tadano,~T.; Tsuneyuki,~S. Self-consistent phonon calculations of lattice dynamical properties in cubic SrTiO 3 with first-principles anharmonic force constants. \emph{Phys. Rev. B} \textbf{2015}, \emph{92}, 054301\relax
\mciteBstWouldAddEndPuncttrue
\mciteSetBstMidEndSepPunct{\mcitedefaultmidpunct}
{\mcitedefaultendpunct}{\mcitedefaultseppunct}\relax
\EndOfBibitem
\bibitem[Zacharias \latin{et~al.}(2023)Zacharias, Volonakis, Giustino, and Even]{zacharias2023anharmonic}
Zacharias,~M.; Volonakis,~G.; Giustino,~F.; Even,~J. Anharmonic lattice dynamics via the special displacement method. \emph{Phys. Rev. B} \textbf{2023}, \emph{108}, 035155\relax
\mciteBstWouldAddEndPuncttrue
\mciteSetBstMidEndSepPunct{\mcitedefaultmidpunct}
{\mcitedefaultendpunct}{\mcitedefaultseppunct}\relax
\EndOfBibitem
\bibitem[Schiltz \latin{et~al.}(2023)Schiltz, Rappoport, and Mandelshtam]{schiltz2023implementation}
Schiltz,~C.; Rappoport,~D.; Mandelshtam,~V.~A. Implementation of the self-consistent phonons method with ab initio potentials (AI-SCP). \emph{J. Chem. Phys.} \textbf{2023}, \emph{158}\relax
\mciteBstWouldAddEndPuncttrue
\mciteSetBstMidEndSepPunct{\mcitedefaultmidpunct}
{\mcitedefaultendpunct}{\mcitedefaultseppunct}\relax
\EndOfBibitem
\bibitem[Monacelli(2024)]{monacelli2024simulating}
Monacelli,~L. Simulating anharmonic crystals: Lights and shadows of first-principles approaches. \emph{arXiv preprint arXiv:2407.03090} \textbf{2024}, \relax
\mciteBstWouldAddEndPunctfalse
\mciteSetBstMidEndSepPunct{\mcitedefaultmidpunct}
{}{\mcitedefaultseppunct}\relax
\EndOfBibitem
\bibitem[Tadano and Tsuneyuki(2019)Tadano, and Tsuneyuki]{tadano2019ab}
Tadano,~T.; Tsuneyuki,~S. Ab initio prediction of structural phase-transition temperature of SrTiO3 from finite-temperature phonon calculation. \emph{J. Ceram. Soc. Jpn.} \textbf{2019}, \emph{127}, 404--408\relax
\mciteBstWouldAddEndPuncttrue
\mciteSetBstMidEndSepPunct{\mcitedefaultmidpunct}
{\mcitedefaultendpunct}{\mcitedefaultseppunct}\relax
\EndOfBibitem
\bibitem[Hoffmann \latin{et~al.}(2019)Hoffmann, Fengler, Herzig, Mittmann, Max, Schroeder, Negrea, Lucian, Slesazeck, and Mikolajick]{hoffmann2019unveiling}
Hoffmann,~M.; Fengler,~F.~P.; Herzig,~M.; Mittmann,~T.; Max,~B.; Schroeder,~U.; Negrea,~R.; Lucian,~P.; Slesazeck,~S.; Mikolajick,~T. Unveiling the double-well energy landscape in a ferroelectric layer. \emph{Nature} \textbf{2019}, \emph{565}, 464--467\relax
\mciteBstWouldAddEndPuncttrue
\mciteSetBstMidEndSepPunct{\mcitedefaultmidpunct}
{\mcitedefaultendpunct}{\mcitedefaultseppunct}\relax
\EndOfBibitem
\bibitem[Konwent(1986)]{konwent1986application}
Konwent,~H. On the Application of a New Type Double-Well Potential in the Theory of Ferroelectric Phase Transitions. \emph{Phys. Status Solidi B} \textbf{1986}, \emph{138}, K7--K11\relax
\mciteBstWouldAddEndPuncttrue
\mciteSetBstMidEndSepPunct{\mcitedefaultmidpunct}
{\mcitedefaultendpunct}{\mcitedefaultseppunct}\relax
\EndOfBibitem
\bibitem[Choudhury \latin{et~al.}(2003)Choudhury, Chitra, and Ramanadham]{choudhury2003role}
Choudhury,~R.~R.; Chitra,~R.; Ramanadham,~M. The role of the double-well potential seen by the amino group in the ferroelectric phase transition in triglycine sulfate. \emph{J. Phys. Condens. Matter} \textbf{2003}, \emph{15}, 4641\relax
\mciteBstWouldAddEndPuncttrue
\mciteSetBstMidEndSepPunct{\mcitedefaultmidpunct}
{\mcitedefaultendpunct}{\mcitedefaultseppunct}\relax
\EndOfBibitem
\bibitem[Wang \latin{et~al.}(2002)Wang, Arago, Garcia, and Gonzalo]{wang2002quantum}
Wang,~C.; Arago,~C.; Garcia,~J.; Gonzalo,~J. Quantum tunneling versus zero-point energy in double-well potential model for ferrroelectric phase transitions. \emph{Phys. A: Stat. Mech. Appl.} \textbf{2002}, \emph{308}, 337--345\relax
\mciteBstWouldAddEndPuncttrue
\mciteSetBstMidEndSepPunct{\mcitedefaultmidpunct}
{\mcitedefaultendpunct}{\mcitedefaultseppunct}\relax
\EndOfBibitem
\bibitem[Fillaux \latin{et~al.}(1998)Fillaux, Nicolai, Baron, Lautie, Tomkinson, and Kearley]{fillaux1998new}
Fillaux,~F.; Nicolai,~B.; Baron,~M.; Lautie,~A.; Tomkinson,~J.; Kearley,~G. A new view of the quantum dynamics for proton transfer along hydrogen bonds: Vibrational spectroscopy with neutrons. \emph{Ber. Bunsenges. Phys. Chem.} \textbf{1998}, \emph{102}, 384--392\relax
\mciteBstWouldAddEndPuncttrue
\mciteSetBstMidEndSepPunct{\mcitedefaultmidpunct}
{\mcitedefaultendpunct}{\mcitedefaultseppunct}\relax
\EndOfBibitem
\bibitem[Fillaux(2002)]{fillaux2002impact}
Fillaux,~F. The impact of vibrational spectroscopy with neutrons on our view of quantum dynamics in hydrogen bonds and proton transfer. \emph{J. Mol. Struct.} \textbf{2002}, \emph{615}, 45--59\relax
\mciteBstWouldAddEndPuncttrue
\mciteSetBstMidEndSepPunct{\mcitedefaultmidpunct}
{\mcitedefaultendpunct}{\mcitedefaultseppunct}\relax
\EndOfBibitem
\bibitem[Xu and Meuwly(2017)Xu, and Meuwly]{xu2017vibrational}
Xu,~Z.-H.; Meuwly,~M. Vibrational spectroscopy and proton transfer dynamics in protonated oxalate. \emph{J. Phys. Chem. A} \textbf{2017}, \emph{121}, 5389--5398\relax
\mciteBstWouldAddEndPuncttrue
\mciteSetBstMidEndSepPunct{\mcitedefaultmidpunct}
{\mcitedefaultendpunct}{\mcitedefaultseppunct}\relax
\EndOfBibitem
\bibitem[Krasilnikov(2014)]{krasilnikov2014two}
Krasilnikov,~P. Two-dimensional model of a double-well potential: Proton transfer upon hydrogen bond deformation. \emph{Biophys.} \textbf{2014}, \emph{59}, 189--198\relax
\mciteBstWouldAddEndPuncttrue
\mciteSetBstMidEndSepPunct{\mcitedefaultmidpunct}
{\mcitedefaultendpunct}{\mcitedefaultseppunct}\relax
\EndOfBibitem
\bibitem[Eckold \latin{et~al.}(1992)Eckold, Grimm, and Stein-Arsic]{eckold1992proton}
Eckold,~G.; Grimm,~H.; Stein-Arsic,~M. Proton disorder and phase transition in KHCO3. \emph{Physica B: Condensed Matter} \textbf{1992}, \emph{180}, 336--338\relax
\mciteBstWouldAddEndPuncttrue
\mciteSetBstMidEndSepPunct{\mcitedefaultmidpunct}
{\mcitedefaultendpunct}{\mcitedefaultseppunct}\relax
\EndOfBibitem
\bibitem[Cailleau \latin{et~al.}(1980)Cailleau, Baudour, Meinnel, Dworkin, Moussa, and Zeyen]{cailleau1980double}
Cailleau,~H.; Baudour,~J.-L.; Meinnel,~J.; Dworkin,~A.; Moussa,~F.; Zeyen,~C.~M. Double-well potentials and structural phase transitions in polyphenyls. \emph{Faraday Discuss.} \textbf{1980}, \emph{69}, 7--18\relax
\mciteBstWouldAddEndPuncttrue
\mciteSetBstMidEndSepPunct{\mcitedefaultmidpunct}
{\mcitedefaultendpunct}{\mcitedefaultseppunct}\relax
\EndOfBibitem
\bibitem[Goryainov(2012)]{goryainov2012model}
Goryainov,~S. A model of phase transitions in double-well Morse potential: Application to hydrogen bond. \emph{Physica B: Condensed Matter} \textbf{2012}, \emph{407}, 4233--4237\relax
\mciteBstWouldAddEndPuncttrue
\mciteSetBstMidEndSepPunct{\mcitedefaultmidpunct}
{\mcitedefaultendpunct}{\mcitedefaultseppunct}\relax
\EndOfBibitem
\bibitem[Siciliano \latin{et~al.}(2024)Siciliano, Monacelli, and Mauri]{siciliano2024beyond}
Siciliano,~A.; Monacelli,~L.; Mauri,~F. Beyond Gaussian fluctuations of quantum anharmonic nuclei: The case of rotational degrees of freedom. \emph{Phys. Rev. B} \textbf{2024}, \emph{110}, 144101\relax
\mciteBstWouldAddEndPuncttrue
\mciteSetBstMidEndSepPunct{\mcitedefaultmidpunct}
{\mcitedefaultendpunct}{\mcitedefaultseppunct}\relax
\EndOfBibitem
\bibitem[Erba \latin{et~al.}(2023)Erba, Desmarais, Casassa, Civalleri, Don\'a, Bush, Searle, Maschio, Edith-Daga, Cossard, Ribaldone, Ascrizzi, Marana, Flament, and Kirtman]{erba2022crystal23}
Erba,~A.; Desmarais,~J.~K.; Casassa,~S.; Civalleri,~B.; Don\'a,~L.; Bush,~I.~J.; Searle,~B.; Maschio,~L.; Edith-Daga,~L.; Cossard,~A.; Ribaldone,~C.; Ascrizzi,~E.; Marana,~N.~L.; Flament,~J.-P.; Kirtman,~B. CRYSTAL23: A Program for Computational Solid State Physics and Chemistry. \emph{J. Chem. Theor. Comput.} \textbf{2023}, \emph{19}, 6891--6932\relax
\mciteBstWouldAddEndPuncttrue
\mciteSetBstMidEndSepPunct{\mcitedefaultmidpunct}
{\mcitedefaultendpunct}{\mcitedefaultseppunct}\relax
\EndOfBibitem
\bibitem[Erba \latin{et~al.}(2019)Erba, Maul, Ferrabone, Carbonni\'ere, R\'erat, and Dovesi]{PARTI_ANHARM}
Erba,~A.; Maul,~J.; Ferrabone,~M.; Carbonni\'ere,~P.; R\'erat,~M.; Dovesi,~R. Anharmonic Vibrational States of Solids from DFT Calculations. Part I: Description of the Potential Energy Surface. \emph{J. Chem. Theory Comput.} \textbf{2019}, \emph{15}, 3755--3765\relax
\mciteBstWouldAddEndPuncttrue
\mciteSetBstMidEndSepPunct{\mcitedefaultmidpunct}
{\mcitedefaultendpunct}{\mcitedefaultseppunct}\relax
\EndOfBibitem
\bibitem[Mitoli \latin{et~al.}(2023)Mitoli, Maul, and Erba]{mitoli2023anharmonic}
Mitoli,~D.; Maul,~J.; Erba,~A. Anharmonic Terms of the Potential Energy Surface: A Group Theoretical Approach. \emph{Cryst. Growth Des.} \textbf{2023}, \emph{23}, 3671--3680\relax
\mciteBstWouldAddEndPuncttrue
\mciteSetBstMidEndSepPunct{\mcitedefaultmidpunct}
{\mcitedefaultendpunct}{\mcitedefaultseppunct}\relax
\EndOfBibitem
\bibitem[Erba \latin{et~al.}(2019)Erba, Maul, Ferrabone, Dovesi, R\'erat, and Carbonni\`ere]{PARTII_ANHARM}
Erba,~A.; Maul,~J.; Ferrabone,~M.; Dovesi,~R.; R\'erat,~M.; Carbonni\`ere,~P. Anharmonic Vibrational States of Solids from DFT Calculations. Part II: Implementation of the VSCF and VCI Methods. \emph{J. Chem. Theor. Comput.} \textbf{2019}, \emph{15}, 3766--3777\relax
\mciteBstWouldAddEndPuncttrue
\mciteSetBstMidEndSepPunct{\mcitedefaultmidpunct}
{\mcitedefaultendpunct}{\mcitedefaultseppunct}\relax
\EndOfBibitem
\bibitem[Maul \latin{et~al.}(2019)Maul, Spoto, Mino, and Erba]{maul2019elucidating}
Maul,~J.; Spoto,~G.; Mino,~L.; Erba,~A. Elucidating the structure and dynamics of CO ad-layers on MgO surfaces. \emph{Physical Chemistry Chemical Physics} \textbf{2019}, \emph{21}, 26279--26283\relax
\mciteBstWouldAddEndPuncttrue
\mciteSetBstMidEndSepPunct{\mcitedefaultmidpunct}
{\mcitedefaultendpunct}{\mcitedefaultseppunct}\relax
\EndOfBibitem
\bibitem[Schireman \latin{et~al.}(2022)Schireman, Maul, Erba, and Ruggiero]{schireman2022anharmonic}
Schireman,~R.~G.; Maul,~J.; Erba,~A.; Ruggiero,~M.~T. Anharmonic Coupling of Stretching Vibrations in Ice: A Periodic VSCF and VCI Description. \emph{J. Chem. Theory Comput.} \textbf{2022}, \emph{18}, 4428--4437\relax
\mciteBstWouldAddEndPuncttrue
\mciteSetBstMidEndSepPunct{\mcitedefaultmidpunct}
{\mcitedefaultendpunct}{\mcitedefaultseppunct}\relax
\EndOfBibitem
\bibitem[Carbonnière \latin{et~al.}(2020)Carbonnière, Erba, Richter, Dovesi, and R{\'e}rat]{carbonniere2020calculation}
Carbonni{\'e}re,~P.; Erba,~A.; Richter,~F.; Dovesi,~R.; R{\'e}rat,~M. Calculation of anharmonic IR and Raman intensities for periodic systems from DFT calculations: Implementation and validation. \emph{J. Chem. Theory Comput.} \textbf{2020}, \emph{16}, 3343--3351\relax
\mciteBstWouldAddEndPuncttrue
\mciteSetBstMidEndSepPunct{\mcitedefaultmidpunct}
{\mcitedefaultendpunct}{\mcitedefaultseppunct}\relax
\EndOfBibitem
\bibitem[Mitoli \latin{et~al.}(2024)Mitoli, Maul, and Erba]{mitoli2024first}
Mitoli,~D.; Maul,~J.; Erba,~A. First-Principles Anharmonic Infrared and Raman Vibrational Spectra of Materials: Fermi Resonance in Dry Ice. \emph{J. Phys. Chem. Lett.} \textbf{2024}, \emph{15}, 888--894\relax
\mciteBstWouldAddEndPuncttrue
\mciteSetBstMidEndSepPunct{\mcitedefaultmidpunct}
{\mcitedefaultendpunct}{\mcitedefaultseppunct}\relax
\EndOfBibitem
\bibitem[Lin \latin{et~al.}(2008)Lin, Gilbert, and Gill]{lin2008calculating}
Lin,~C.~Y.; Gilbert,~A.~T.; Gill,~P.~M. Calculating molecular vibrational spectra beyond the harmonic approximation. \emph{Theor. Chem. Acc.} \textbf{2008}, \emph{120}, 23--35\relax
\mciteBstWouldAddEndPuncttrue
\mciteSetBstMidEndSepPunct{\mcitedefaultmidpunct}
{\mcitedefaultendpunct}{\mcitedefaultseppunct}\relax
\EndOfBibitem
\bibitem[Balsa \latin{et~al.}(1983)Balsa, Plo, Esteve, and Pacheco]{balsa1983simple}
Balsa,~R.; Plo,~M.; Esteve,~J.; Pacheco,~A. Simple procedure to compute accurate energy levels of a double-well anharmonic oscillator. \emph{Phys. Rev. D} \textbf{1983}, \emph{28}, 1945\relax
\mciteBstWouldAddEndPuncttrue
\mciteSetBstMidEndSepPunct{\mcitedefaultmidpunct}
{\mcitedefaultendpunct}{\mcitedefaultseppunct}\relax
\EndOfBibitem
\bibitem[Camino \latin{et~al.}(2023)Camino, Zhou, Ascrizzi, Boccuni, Bodo, Cossard, Mitoli, Ferrari, Erba, and Harrison]{CRYSTALpytools}
Camino,~B.; Zhou,~H.; Ascrizzi,~E.; Boccuni,~A.; Bodo,~F.; Cossard,~A.; Mitoli,~D.; Ferrari,~A.~M.; Erba,~A.; Harrison,~N.~M. CRYSTALpytools: a Python Infrastructure for the CRYSTAL Code. \emph{Comput. Phys. Commun.} \textbf{2023}, \emph{292}, 108853\relax
\mciteBstWouldAddEndPuncttrue
\mciteSetBstMidEndSepPunct{\mcitedefaultmidpunct}
{\mcitedefaultendpunct}{\mcitedefaultseppunct}\relax
\EndOfBibitem
\bibitem[McKenzie and Dryden(1973)McKenzie, and Dryden]{mckenzie1973dielectric}
McKenzie,~D.; Dryden,~J. Dielectric properties and ferroelectric transitions of thiourea. \emph{J. Phys. C: Solid State Phys.} \textbf{1973}, \emph{6}, 767\relax
\mciteBstWouldAddEndPuncttrue
\mciteSetBstMidEndSepPunct{\mcitedefaultmidpunct}
{\mcitedefaultendpunct}{\mcitedefaultseppunct}\relax
\EndOfBibitem
\bibitem[Klimowski \latin{et~al.}(1976)Klimowski, Wanarski, and O{\.z}go]{klimowski1976effect}
Klimowski,~J.; Wanarski,~W.; O{\.z}go,~D. Effect of hydrostatic pressure on the phase transition temperatures and spontaneous polarization of thiourea monocrystals. \emph{Phys. Status Solidi A} \textbf{1976}, \emph{34}, 697--704\relax
\mciteBstWouldAddEndPuncttrue
\mciteSetBstMidEndSepPunct{\mcitedefaultmidpunct}
{\mcitedefaultendpunct}{\mcitedefaultseppunct}\relax
\EndOfBibitem
\bibitem[Dove and Lynden-bell(1986)Dove, and Lynden-bell]{dove1986model}
Dove,~M.~T.; Lynden-bell,~R.~M. A model of the paraelectric phase of thiourea. \emph{Phylos. Mag. B} \textbf{1986}, \emph{54}, 443--463\relax
\mciteBstWouldAddEndPuncttrue
\mciteSetBstMidEndSepPunct{\mcitedefaultmidpunct}
{\mcitedefaultendpunct}{\mcitedefaultseppunct}\relax
\EndOfBibitem
\bibitem[Elcombe and Taylor(1968)Elcombe, and Taylor]{elcombe1968neutron}
Elcombe,~M.~M.; Taylor,~J. A neutron diffraction determination of the crystal structures of thiourea and deuterated thiourea above and below the ferroelectric transition. \emph{Acta Crystallogr. A} \textbf{1968}, \emph{24}, 410--420\relax
\mciteBstWouldAddEndPuncttrue
\mciteSetBstMidEndSepPunct{\mcitedefaultmidpunct}
{\mcitedefaultendpunct}{\mcitedefaultseppunct}\relax
\EndOfBibitem
\bibitem[Siapkas(1980)]{siapkas1980soft}
Siapkas,~D. Soft mode in the disordered-incommensurate-commensurate phase transitions of thiourea. \emph{Ferroelectrics} \textbf{1980}, \emph{29}, 29--32\relax
\mciteBstWouldAddEndPuncttrue
\mciteSetBstMidEndSepPunct{\mcitedefaultmidpunct}
{\mcitedefaultendpunct}{\mcitedefaultseppunct}\relax
\EndOfBibitem
\bibitem[Petzelt(1981)]{petzelt1981dielectric}
Petzelt,~J. Dielectric and light scattering spectroscopy of incommensurate phases in crystals. \emph{Ph. Transit.} \textbf{1981}, \emph{2}, 155--229\relax
\mciteBstWouldAddEndPuncttrue
\mciteSetBstMidEndSepPunct{\mcitedefaultmidpunct}
{\mcitedefaultendpunct}{\mcitedefaultseppunct}\relax
\EndOfBibitem
\bibitem[Erba \latin{et~al.}(2017)Erba, Baima, Bush, Orlando, and Dovesi]{MPP2017}
Erba,~A.; Baima,~J.; Bush,~I.; Orlando,~R.; Dovesi,~R. Large Scale Condensed Matter DFT Simulations: Performance and Capabilities of the \textsc{Crystal} Code. \emph{J. Chem. Theory Comput.} \textbf{2017}, \emph{13}, 5019--5027\relax
\mciteBstWouldAddEndPuncttrue
\mciteSetBstMidEndSepPunct{\mcitedefaultmidpunct}
{\mcitedefaultendpunct}{\mcitedefaultseppunct}\relax
\EndOfBibitem
\bibitem[Becke(1993)]{B3LYP}
Becke,~A.~D. Density-functional thermochemistry. III. The role of exact exchange. \emph{J. Chem. Phys.} \textbf{1993}, \emph{98}, 5648\relax
\mciteBstWouldAddEndPuncttrue
\mciteSetBstMidEndSepPunct{\mcitedefaultmidpunct}
{\mcitedefaultendpunct}{\mcitedefaultseppunct}\relax
\EndOfBibitem
\bibitem[Grimme \latin{et~al.}(2010)Grimme, Antony, Ehrlich, and Krieg]{GRIMMED3}
Grimme,~S.; Antony,~J.; Ehrlich,~S.; Krieg,~H. A consistent and accurate ab initio parametrization of density functional dispersion correction (DFT-D) for the 94 elements H-Pu. \emph{J. Chem. Phys.} \textbf{2010}, \emph{132}, 154104\relax
\mciteBstWouldAddEndPuncttrue
\mciteSetBstMidEndSepPunct{\mcitedefaultmidpunct}
{\mcitedefaultendpunct}{\mcitedefaultseppunct}\relax
\EndOfBibitem
\bibitem[Vilela~Oliveira \latin{et~al.}(2019)Vilela~Oliveira, Laun, Peintinger, and Bredow]{vilela2019bsse}
Vilela~Oliveira,~D.; Laun,~J.; Peintinger,~M.~F.; Bredow,~T. BSSE-correction scheme for consistent gaussian basis sets of double-and triple-zeta valence with polarization quality for solid-state calculations. \emph{J. Comput. Chem.} \textbf{2019}, \emph{40}, 2364--2376\relax
\mciteBstWouldAddEndPuncttrue
\mciteSetBstMidEndSepPunct{\mcitedefaultmidpunct}
{\mcitedefaultendpunct}{\mcitedefaultseppunct}\relax
\EndOfBibitem
\bibitem[Maschio \latin{et~al.}(2012)Maschio, Kirtman, Orlando, and R{\`e}rat]{maschio2012ab}
Maschio,~L.; Kirtman,~B.; Orlando,~R.; R{\`e}rat,~M. Ab initio analytical infrared intensities for periodic systems through a coupled perturbed Hartree-Fock/Kohn-Sham method. \emph{J. Chem. Phys.} \textbf{2012}, \emph{137}\relax
\mciteBstWouldAddEndPuncttrue
\mciteSetBstMidEndSepPunct{\mcitedefaultmidpunct}
{\mcitedefaultendpunct}{\mcitedefaultseppunct}\relax
\EndOfBibitem
\bibitem[Dovesi \latin{et~al.}(2018)Dovesi, Kirtman, Maschio, Maul, Pascale, and R{\'e}rat]{dovesi2018calculation}
Dovesi,~R.; Kirtman,~B.; Maschio,~L.; Maul,~J.; Pascale,~F.; R{\'e}rat,~M. Calculation of the infrared intensity of crystalline systems. A comparison of three strategies based on berry phase, wannier function, and coupled-perturbed Kohn--Sham methods. \emph{J. Phys. Chem. C} \textbf{2018}, \emph{123}, 8336--8346\relax
\mciteBstWouldAddEndPuncttrue
\mciteSetBstMidEndSepPunct{\mcitedefaultmidpunct}
{\mcitedefaultendpunct}{\mcitedefaultseppunct}\relax
\EndOfBibitem
\end{mcitethebibliography}

\end{document}